# BLOCKCHAIN AS A PLATFORM FOR ARTIFICIAL INTELLIGENCE (AI) TRANSPARENCY

Afroja Akther[1], Ayesha Arobee[2], Abdullah Al Adnan[3], Omum Auyon[4], ASM Johirul Islam[5], Farhad Akter[6]

[123456]Emporia State University, Kansas, USA
[1]Corresponding author: afrojabnkdu@gmail.com

## ABSTRACT

*As artificial intelligence (AI) systems become increasingly complex and autonomous, concerns over transparency and accountability have intensified. The "black box" problem in AI decision-making limits stakeholders' ability to understand, trust, and verify outcomes, particularly in high-stakes sectors such as healthcare, finance, and autonomous systems. Blockchain technology, with its decentralized, immutable, and transparent characteristics, presents a potential solution to enhance AI transparency and auditability. This paper explores the integration of blockchain with AI to improve decision traceability, data provenance, and model accountability. By leveraging blockchain as an immutable record-keeping system, AI decision-making can become more interpretable, fostering trust among users and regulatory compliance. However, challenges such as scalability, integration complexity, and computational overhead must be addressed to fully realize this synergy. This study discusses existing research, proposes a framework for blockchain-enhanced AI transparency, and highlights practical applications, benefits, and limitations. The findings suggest that blockchain could be a foundational technology for ensuring AI systems remain accountable, ethical, and aligned with regulatory standards.*

## KEYWORDS

*Artificial Intelligence (AI), blockchain, ai transparency, black box problem, decentralization*

## 1. INTRODUCTION

Artificial intelligence (AI) continues to transform various sectors, including healthcare, finance, transportation, driving innovation and efficiency. However, as AI systems become more complex and autonomous, their decision-making processes often lack transparency, leading to what is commonly known as the "black box" problem. This opacity undermines trust among users and stakeholders, posing a significant barrier to the broader adoption of AI technologies. The trust deficit is particularly pronounced in high-stakes sectors, where AI-driven decisions can have profound ethical, legal, and safety implications [1]. Transparency in AI refers to the clarity and interpretability of decision-making processes within AI systems. It is crucial for validation, debugging, and building trust, both from an ethical standpoint and for practical implementation [2]. Transparent AI systems allow users to understand and challenge decisions, fostering accountability and compliance with regulatory standards such as the General Data Protection Regulation (GDPR) [3]. Despite advancements in explainable AI (XAI), many AI models, particularly deep learning systems, remain opaque, making it difficult to trace or justify their decisions.

Blockchain technology offers unique features that can address AI transparency challenges. Known for its core attributes—decentralization, immutability, and transparency—blockchain provides a structured way to record and trace the decision-making process in AI systems [4]. Each transaction recorded on a blockchain is time-stamped and cryptographically linked to previous transactions, creating an auditable trail that enhances security and trustworthiness [5]. By integrating blockchain with AI, organizations can ensure that every decision, data input, and model update is permanently recorded, verifiable, and accessible to auditors, regulators, and the public [6]. This approach can revolutionize how trust and transparency are perceived in AI applications, particularly in finance, healthcare, and autonomous systems, where decisions must be explainable and accountable [7].

This research aims to explore the potential of blockchain technology in enhancing AI transparency, auditability, and trust. Specifically, it seeks to answer the following research questions -

   a. How can blockchain be leveraged to enhance AI transparency and explainability?

   b. What are the key challenges and limitations of integrating blockchain with AI?

   c. How can blockchain-based AI transparency frameworks align with existing regulatory and ethical standards?

By documenting the development, deployment, and operational phases of AI within a blockchain framework, this study evaluates whether such integration can make AI decision-making processes more understandable and acceptable to a broader audience.

## 2. THE CHALLENGE OF AI TRANSPARENCY

Artificial intelligence (AI), particularly in its more sophisticated forms such as deep learning, has introduced unprecedented capabilities in data analysis and decision-making. However, these advances have also brought significant challenges, chiefly the opacity of the decision-making processes. This opacity, often referred to as the "black box" problem, describes situations where the internal workings of AI systems are not visible or understandable to users or stakeholders [8]. The concern is that if users cannot see or understand how decisions are made, they cannot verify the fairness, accuracy, or safety of these decisions.

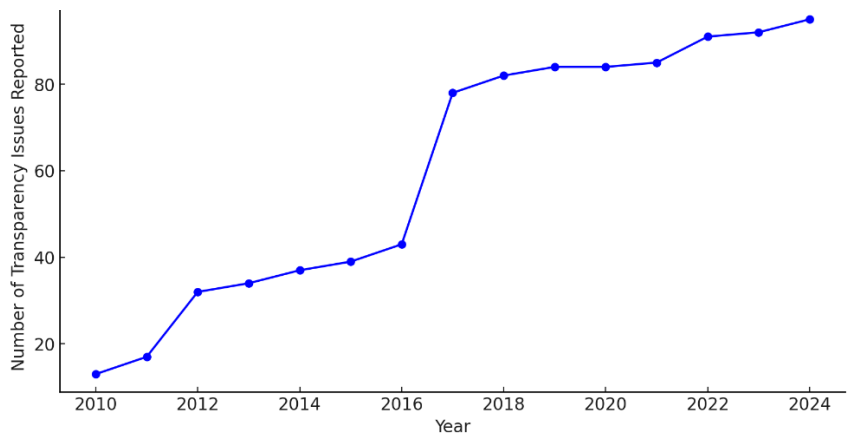

Figure 1. The Challenges of AI Transparency

The necessity for transparency (Figure 1) becomes crucial in sectors where decisions have significant consequences –

*Healthcare: I*n healthcare, AI systems are used for diagnosing diseases, recommending treatments, and managing patient data. The accuracy and explicability of AI decisions can directly affect patient outcomes and safety [9].

*Finance:* In finance, AI determines credit ratings, manages investments, and detects fraud. Decisions need to be transparent to ensure fairness, avoid biases, and comply with regulations like the General Data Protection Regulation (GDPR), which includes rights to explanation [3].

*Autonomous Driving:* For autonomous vehicles, AI systems make real-time decisions about navigation and safety. Transparency is essential to build user trust and facilitate regulatory approval [10] (McAllister et al., 2017).

Table 1: Sectors Impacted by AI Transparency Challenges

| Sector | Impact of AI Transparency | Citations |
| --- | --- | --- |
| Healthcare | In healthcare, AI's role in diagnosing diseases, recommending treatments, and managing patient data demands high transparency to ensure patient safety and treatment efficacy. | [9] |
| Finance | AI systems in finance determine credit ratings, manage investments, and detect fraud. Transparency is crucial to ensure fairness, avoid biases, and comply with regulations. | [3] |
| Autonomous Driving | For autonomous vehicles, transparent AI decision-making is essential for safety, user trust, and regulatory approval. | [10] |
| Legal | AI applications in the legal sector, such as predictive policing and recidivism risk assessment, require transparency to prevent biases and uphold legal standards and fairness. | [11] |
| Employment | In employment, AI tools are used for screening resumes and automating hiring decisions. Transparency helps ensure non-discriminatory practices and adherence to employment laws. | [12] |
| Retail | AI in retail enhances customer experience through personalized recommendations and inventory management. Transparency is needed to protect consumer data and ensure privacy. | [13] |

The primary technical challenge in enhancing transparency is the inherent complexity of machine learning models, especially deep neural networks. These models involve large numbers of parameters and layers, which makes understanding their function and predicting their behavior difficult. This complexity is compounded by the stochastic nature of many AI algorithms, where even slight changes in input data or initial conditions can significantly alter outcomes. Table 2 reveals several approaches have been proposed to tackle the transparency issue in AI.

Table 2: Strategies for Improving AI Transparency

| Strategy | Description | References |
| --- | --- | --- |
| Simplification of Models | Using simpler models such as decision trees or linear regression that are more interpretable compared to complex models like neural networks. These are easier to understand but may offer less analytical power. | [14] |
| Development of | Tools like LIME and SHAP have been developed to make the output of complex models more interpretable to users, allowing for better understanding of how decisions are made. | [15] |
| Interpretability Tools | | |
| Incorporating Transparency by Design | Embedding transparency into the AI development process to ensure systems are understandable from the beginning, rather than attempting to add explanations to pre-existing models. | [16] |

Transparency is not only a technical requirement but also an ethical imperative. Ethical guidelines for AI, such as those proposed by the EU's High-Level Expert Group on Artificial Intelligence, emphasize respect for human autonomy, prevention of harm, fairness, and explicability as key requirements [17]. Regulatory frameworks are also evolving to address these challenges, emphasizing the need for transparency to ensure that AI systems do not perpetuate biases or make unjustifiable decisions.

Addressing the transparency challenge in AI is critical for its ethical application and broader acceptance, especially in high-stakes domains. Ongoing research into interpretability, the development of new tools, and the creation of robust regulatory frameworks are essential to ensuring that AI systems operate transparently and are accountable for their decisions.

## 3. BLOCKCHAIN FUNDAMENTALS

Blockchain technology, initially popularized by the cryptocurrency Bitcoin, is fundamentally a distributed ledger that records transactions across multiple computers in such a way that the registered transactions cannot be altered retroactively [18]. This technology provides a robust framework for facilitating transactions without the need for a trusted third party, such as a bank or regulatory agency [5] [19]. This section explores the core characteristics of blockchain technology—decentralization, immutability, and transparency—and their implications for enhancing AI transparency. Decentralization refers to the distribution of control and authority across all network participants. Unlike traditional centralized systems where a single entity has control, blockchain distributes the control to all participating nodes in the network. This not only eliminates a single point of failure but also increases resistance to malicious activities [20]. In the context of AI, decentralization can democratize data handling, allowing multiple stakeholders to participate in training and decision-making processes of AI models. This is particularly beneficial in scenarios like federated learning, where multiple entities collaborate to improve a model while maintaining control over their own data.

Immutability in blockchain refers to the characteristic that, once a transaction has been recorded in the distributed ledger, it cannot be altered or deleted by any single party. This is ensured through cryptographic hash functions and the consensus mechanisms that require majority approval from all nodes in the network for changes to be made [21]. For AI systems, immutability means that every decision, data input, model adjustment, and output can be permanently recorded and verified. This assures stakeholders of the integrity of the data used and the decisions made by AI systems, which is critical in sectors requiring high levels of auditability such as healthcare and financial services.

Transparency in blockchain technology means that all transactions and their details are visible to anyone who has access to the network. While individual user identities can be protected through pseudonyms, the transactions themselves are open to scrutiny [22]. Transparency helps in making AI systems more understandable and accountable. Stakeholders can trace the decision-making process, understand the data inputs and outputs, and verify the actions taken by an AI system. This level of openness is vital for building trust, particularly in applications affecting public services or consumer rights.

### 3.1 Practical Applications and Challenges

Blockchain's integration with AI has practical applications in numerous fields. For example, in supply chain management, blockchain can provide a transparent, immutable record of product provenance and handling, which can be combined with AI to optimize logistics and predict supply chain disruptions [23]. Despite its potential, blockchain faces challenges such as scalability, energy consumption, and complexity integration with existing technologies. These challenges need to be addressed to fully harness the power of blockchain in AI transparency [24]. The fundamental characteristics of blockchain—decentralization, immutability, and transparency—offer a strong foundation for addressing the transparency issues in AI systems. By embedding these features into AI workflows, stakeholders can achieve a higher degree of trust and accountability, paving the way for more responsible and acceptable AI applications.

## 4. INTEGRATING BLOCKCHAIN WITH AI

The integration of blockchain technology with artificial intelligence (AI) systems (Table 3) presents a transformative approach to enhancing transparency, accountability, and trust in AI operations. Blockchain's core features—decentralization, immutability, and transparency—can significantly augment AI functionalities in various ways, from decision-making processes to data handling and model management.

### 4.1 Tracking and Recording Decisions

Every decision made by an AI system, from simple classifications to complex predictive analytics, can be recorded on a blockchain. This recording process includes the inputs, decision-making parameters, and outputs of the AI system. Such a transparent and immutable record ensures that every decision is traceable and auditable. This feature is particularly valuable in environments where decisions must be explainable and verifiable, such as in regulatory compliance or legal investigations [25]. Implementing this would require linking AI decision-making processes directly to a blockchain interface, where each decision triggers a transaction on the blockchain. This could be facilitated through smart contracts that automatically execute and record transactions when certain conditions are met (Figure 2).

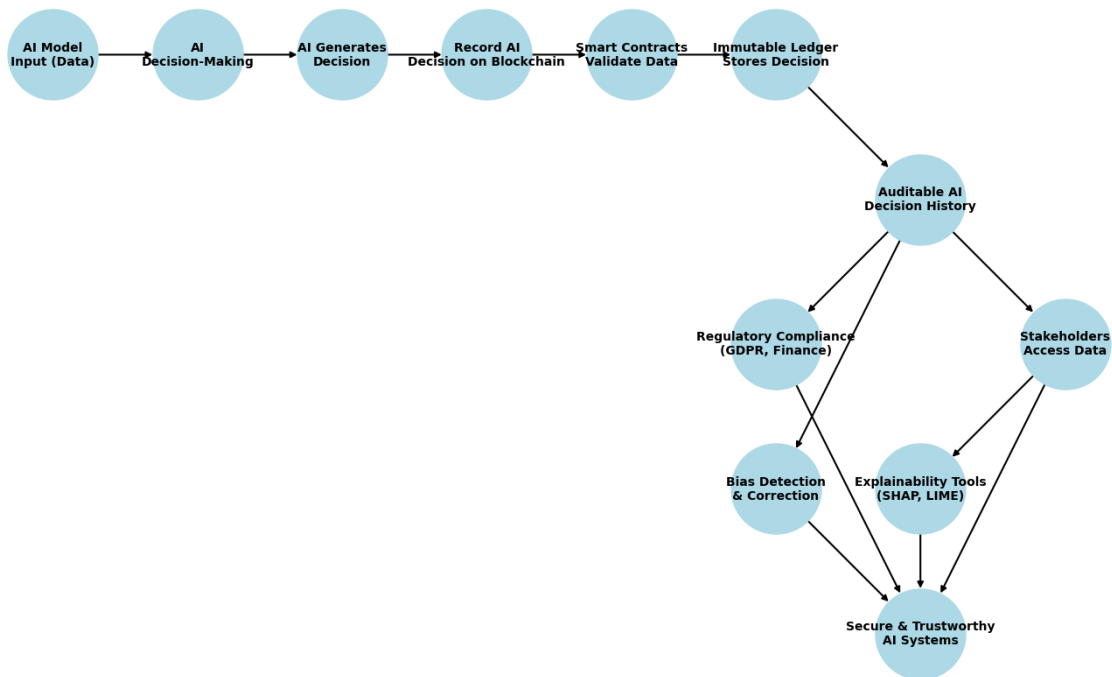

Figure 2. Blockchain enhanced AI transparency framework (Source: Authors)

### 4.2 Considerations for Blockchain Architectures and AI Model Variability

While blockchain technology offers promising solutions for enhancing AI transparency, the selection of appropriate blockchain architecture is crucial for optimizing its effectiveness. Existing blockchain frameworks, such as Ethereum, Hyperledger Fabric, and Corda, each have unique characteristics that impact their suitability for recording AI decision-making processes. Ethereum, with its smart contract capabilities, provides a decentralized and programmable infrastructure, while Hyperledger Fabric offers permissioned ledger functionality, which may be preferable in enterprise or regulatory environments. Corda, designed for financial applications, emphasizes privacy and efficient transaction validation. A comparative analysis of these architectures would provide deeper insight into their applicability for AI governance and auditability.

Furthermore, the degree to which blockchain enhances AI transparency depends on the underlying AI model. Neural networks, decision trees, and ensemble learning models exhibit varying levels of interpretability. Decision trees, for instance, are inherently transparent due to their rule-based structure, while deep neural networks (DNNs) function as "black boxes," making their decision-making processes less interpretable. Blockchain's role in ensuring AI transparency may therefore differ depending on the model in question. Integrating blockchain with explainability frameworks such as SHAP (Shapley Additive Explanations), LIME (Local Interpretable Model-agnostic Explanations), or counterfactual reasoning could enhance the traceability and verifiability of AI-driven decisions across different model types. Future research should explore how blockchain's immutable ledger can be effectively leveraged to store model explanations, thereby improving accountability and regulatory compliance in AI applications.

Table 3: Integrating Blockchain with AI

| Integration Aspect | Description | Impacts and Benefits | References |
|---|---|---|---|
| Tracking and Recording Decisions | Every AI decision is recorded on blockchain, including inputs, parameters, and outputs. This process ensures transparency and auditability. | Enhances the explainability and verifiability of AI decisions, crucial in environments requiring regulatory compliance or legal scrutiny. | [25] |
| Data Handling and Usage | Blockchain logs details about the datasets used by AI, including data provenance, usage, and changes. | Ensures data integrity and consistency, critical for applications where data history influences outputs, such as financial forecasting or clinical support systems. | [26] |
| Model Management | Blockchain manages AI models, tracking versions, updates, performance metrics, and deployment records. | Facilitates understanding of AI model lifecycle, aids in identifying which models were used and why, and supports rollback features to previous model versions if needed. | [27] |

### 4.3 Data Handling and Usage

Blockchain can be used to log the datasets utilized by AI systems throughout their learning and operational processes. This includes recording details about data provenance, access, and changes over time. By using blockchain to monitor data handling, stakeholders can ensure the integrity and consistency of data used by AI systems. This is crucial for applications where data sources and history significantly influence the system's outputs, such as in financial forecasting or clinical decision support systems [26]. A blockchain ledger could capture metadata about each dataset accessed by the AI system, including who provided the data, when it was used, and any alterations made. This process not only aids in tracking data lineage but also helps in maintaining compliance with data governance standards and privacy regulations.

### 4.4 Model Management

Blockchain can facilitate the management of AI models, including their versions, updates, performance metrics, and deployment records. This aspect of blockchain integration helps maintain a chronological log of model evolution. Effective model management is critical for understanding the lifecycle of AI models, particularly in complex systems where multiple models are tested and deployed over time. Recording model changes on a blockchain can aid developers and stakeholders in identifying which models were used in specific instances and why changes were made [27]. Blockchain can store hashes of model versions along with detailed descriptions and performance metrics. Each time a model is updated, a new entry is made in the blockchain, creating an immutable history of the model's development and use. This approach can also support rollback features where an earlier model version can be reinstated from the blockchain records if needed.

Integrating blockchain with AI can significantly enhance the transparency, reliability, and auditability of AI systems. This integration not only addresses trust issues associated with AI but also paves the way for new standards in the development and deployment of trustworthy AI systems. By recording every key action and decision on a blockchain, stakeholders can ensure that AI systems are not only powerful and efficient but also accountable and open to scrutiny.

## 5. POTENTIAL BENEFITS OF USING BLOCKCHAIN FOR AI TRANSPARENCY

### 5.1 Increased Trust

Trust is foundational to the broader acceptance and adoption of AI systems, particularly in sectors where decisions have significant implications, such as healthcare, finance, and autonomous systems. Blockchain can enhance trust in AI by providing a transparent, immutable record of all AI operations, including data inputs, decision-making processes, and outputs. This transparency ensures that AI decisions are not only visible but also verifiable, making it easier for users and stakeholders to trust the output generated by AI systems [28]. By allowing stakeholders to audit and validate AI decisions through blockchain records, any anomalies or biases in AI operations can be detected and addressed. This leads to improved trust, which is crucial for industries that rely on public confidence to operate effectively.

### 5.2 Enhanced Security

Security concerns in AI involve unauthorized data access, tampering, and potential misuse of AI technologies. Blockchain addresses these issues with its inherent security features such as cryptographic hash functions, consensus algorithms, and distributed network architecture, which ensure that once a transaction is added to the blockchain, it cannot be altered or deleted by any single entity without detection [29]. Implementing blockchain within AI frameworks can protect against tampering and enhance the overall security of AI systems. For example, in AI-driven supply chain solutions, blockchain can secure data across the network, from the manufacturer to the end consumer, ensuring that all decisions and transactions are accurately recorded and resistant to fraud.

### 5.3 Regulatory Compliance

As AI technologies become integral to more aspects of daily life, governments and regulatory bodies are increasingly focusing on establishing frameworks that ensure these technologies are safe, fair, and accountable. Regulations such as the GDPR in Europe have started to emphasize the need for transparency and accountability in automated decision-making processes [30]. Blockchain can facilitate compliance with these regulatory requirements by providing a mechanism for documenting and verifying every step in the AI decision-making process. This capability is particularly important for meeting regulations that require explainability, such as providing individuals with reasons for decisions made by AI systems that affect them. Blockchain's audit trails enable organizations to demonstrate compliance in a straightforward and verifiable manner, potentially simplifying the regulatory review and auditing processes.

### 5.4 Enhancing the Framework and Practical Applications

While the proposed framework for integrating blockchain with AI transparency outlines key principles, it requires further elaboration to establish a more comprehensive and implementable model. A well-defined architectural framework detailing the interaction between blockchain components (e.g., smart contracts, distributed ledgers, and consensus mechanisms) and AI models (e.g., neural networks, decision trees, and reinforcement learning systems) would provide clearer insights into how blockchain can enhance AI transparency in practical scenarios. Additionally, a conceptual diagram or system architecture illustrating the data flow, decision verification process, and auditability features within a blockchain-integrated AI system would significantly improve the framework's clarity and applicability.

Beyond theoretical discussions, the paper lacks empirical evidence or case studies demonstrating successful implementations of blockchain-enhanced AI transparency. Several real-world applications highlight the potential of this integration -

a. Healthcare – AI-Powered Diagnostics and Blockchain for Auditability: A study by Kuo et al. (2017) explored how blockchain technology can enhance electronic health records (EHRs) by ensuring data integrity and patient consent tracking. When combined with AI-driven diagnostic tools, blockchain can provide an immutable record of diagnostic decisions, allowing medical practitioners to verify how AI-derived conclusions were reached. This approach enhances trust in AI-powered medical decision-making, particularly in critical diagnoses like oncology and radiology [9].

b. Finance – Fraud Detection and Blockchain-Based Transaction Transparency: In financial applications, AI is widely used for fraud detection and risk assessment. Wang et al. (2018) demonstrated that integrating AI with blockchain enhances fraud detection models by ensuring transaction traceability and preventing data manipulation [27]. AI models can analyze transaction patterns, while blockchain serves as a tamper-proof ledger to maintain the integrity of AI's decision-making history, thereby improving compliance with regulations like the General Data Protection Regulation (GDPR) [3].

c. Autonomous Systems – Blockchain for Explainable AI in Self-Driving Vehicles: AI-driven autonomous vehicles rely on deep learning models to process sensor data and make real-time driving decisions. However, McAllister et al. (2017) argue that a lack of transparency in these decisions remains a critical challenge for regulatory approval [10]. By storing decision logs on blockchain, regulators and manufacturers can audit vehicle behavior in accident scenarios, ensuring accountability and legal compliance. Companies such as Bosch and IBM have initiated projects using blockchain to secure autonomous driving data, reinforcing AI's explainability and reducing liability concerns.

While these case studies provide promising insights, further empirical validation is required to measure the performance of blockchain-enhanced AI transparency in diverse sectors. Future research should focus on developing and testing blockchain-AI integration models in real-world environments, assessing their impact on scalability, efficiency, and regulatory compliance. Conducting comparative analyses of different blockchain architectures (e.g., Ethereum, Hyperledger, Corda) within AI-driven decision-making frameworks would offer valuable insights into selecting the most suitable infrastructure for enhancing AI transparency. By incorporating empirical data and practical implementations, this research can better inform policymakers, industry stakeholders, and researchers about the viability of blockchain as a foundational technology for AI governance.

Table 4: Potential Benefits of Using Blockchain for AI Transparency

| Benefit | Explanation and Impact | References |
|---|---|---|
| Increased Trust | Blockchain enhances trust in AI by making AI decisions auditable and understandable, providing a transparent record of AI operations. Impact: This visibility allows users and stakeholders to verify AI decisions, essential for trust, particularly in critical sectors like healthcare and finance. | [28] |
| Enhanced Security | Blockchain's security features, such as cryptographic hashes and consensus algorithms, protect AI systems from tampering and unauthorized access. Impact: This secures data across networks, crucial for AI-driven processes where data integrity is fundamental, such as in supply chains. | [29] |
| Regulatory Compliance | Blockchain supports compliance with regulations that demand transparency and accountability in AI, such as the GDPR. Impact: Blockchain's audit trails help organizations demonstrate compliance effectively, crucial for operations impacted by automated decision-making regulations. | [30] |

Integrating blockchain with AI offers significant benefits (Table 4) that can address some of the key challenges currently faced by AI technologies, particularly in terms of trust, security, and regulatory compliance. The decentralized, immutable, and transparent nature of blockchain makes it an ideal technology to enhance the accountability and trustworthiness of AI systems, thereby facilitating wider adoption and more responsible use of AI across various sectors.

## 6. CHALLENGES AND CONSIDERATIONS IN INTEGRATING BLOCKCHAIN WITH AI

### 6.1 Scalability

Blockchain technology, particularly when based on proof-of-work (PoW) consensus mechanisms, often struggles with scalability issues. The inherent design of many blockchains limits their transaction throughput, which can lead to delays and increased costs for recording transactions [31]. AI systems, especially those requiring real-time data processing and decision-making, need high-speed data access and processing capabilities. The slow transaction speeds and potential bottlenecks of blockchain can significantly hinder the performance of AI applications that depend on large datasets and immediate responses, such as those used in autonomous driving or real-time financial trading [32].

### 6.2 Complexity

Integrating blockchain with AI adds a layer of complexity to both systems. Blockchain networks require maintenance of consensus protocols, execution of smart contracts, and management of cryptographic keys, among other aspects. The added complexity could impact on the overall efficiency and performance of AI systems. For instance, the computational overhead required to maintain blockchain integrity and execute smart contracts can divert resources away from core AI processing tasks, potentially degrading system performance [33].

### 6.3 Technological Integration

Both AI and blockchain are sophisticated technologies that operate on different paradigms. AI involves adaptive learning processes and probabilistic logic, while blockchain is based on deterministic and immutable record-keeping. Bridging the technological gap requires meticulous architectural planning and integration. The challenge lies in ensuring that blockchain does not impede the learning efficiency of AI or its ability to update and adapt dynamically to new data. This integration must be thoughtfully designed to maintain the strengths of both technologies without compromising their core functionalities [34].

### 6.4 Addressing Challenges in Blockchain-AI Integration

While blockchain technology presents a promising approach for enhancing AI transparency, its implementation is not without challenges. The paper acknowledges key limitations such as scalability and integration complexities; however, potential solutions and mitigation strategies for these issues remain underexplored. One primary scalability concern arises from blockchain's consensus mechanisms, particularly Proof-of-Work (PoW), which can be computationally intensive and result in low transaction throughput. Alternative consensus protocols, such as Proof-of-Stake (PoS), Delegated Proof-of-Stake (DPoS), and Directed Acyclic Graphs (DAGs), could mitigate these concerns by improving processing speed and reducing energy consumption. Additionally, layer-2 scaling solutions such as sharding, rollups, and state channels have been proposed to enhance blockchain efficiency without compromising security or decentralization. Future research should evaluate the feasibility of these approaches in AI-integrated blockchain systems.

Another critical challenge pertains to energy consumption, particularly in blockchains utilizing PoW-based mechanisms. The high computational requirements of PoW contribute to significant energy expenditures, raising concerns about sustainability. Exploring energy-efficient blockchain frameworks, such as Hyperledger Fabric, which employs a permissioned and optimized consensus mechanism, or Ethereum's transition to PoS, could provide insights into more sustainable blockchain applications for AI governance. Assessing the trade-offs between decentralization, security, and energy efficiency will be essential for developing AI-integrated blockchain solutions that align with environmental sustainability goals.

Beyond technical concerns, the ethical implications of using blockchain for AI transparency require deeper examination. While blockchain enhances auditability and accountability, it also raises questions regarding data privacy and anonymization. The immutability of blockchain records, while beneficial for maintaining an auditable history, could conflict with privacy regulations such as the General Data Protection Regulation (GDPR), which mandates the right to be forgotten. Solutions such as zero-knowledge proofs (ZKPs), homomorphic encryption, and privacy-preserving smart contracts could help balance transparency with data protection. Additionally, blockchain's role in mitigating AI biases and ensuring fair decision-making processes must be further explored to address ethical concerns surrounding algorithmic accountability and governance.

Future research should focus on developing hybrid models that integrate privacy-preserving blockchain mechanisms with AI explainability frameworks to ensure a balance between transparency, efficiency, and ethical compliance. These considerations will be critical in fostering the responsible and scalable deployment of blockchain-integrated AI systems.

Table 5: Challenges and Considerations in Integrating Blockchain with AI

| Challenge Category | Description | Specific Issues and Impacts | References |
|---|---|---|---|
| Scalability | Blockchain, especially using proof-of-work (PoW) mechanisms, faces scalability issues due to limited transaction throughput. | AI systems requiring real-time data processing find blockchain's slow transaction speeds a significant bottleneck. This is critical in applications needing immediate responses, like autonomous driving or real-time financial trading. | [31] |
| Complexity | Integrating blockchain adds complexity due to the need for maintaining consensus protocols, managing | The additional computational overhead can reduce the efficiency and performance of AI systems. Resources needed to maintain blockchain integrity and execute smart contracts may divert | [33] |

| | cryptographic keys, and executing smart contracts. | from core AI tasks, affecting overall system performance. | |
|---|---|---|---|
| Technological Integration | AI and blockchain operate on fundamentally different paradigms: AI on adaptive learning and probabilistic logic, and blockchain on deterministic and immutable record-keeping. | Bridging the gap between these technologies requires careful planning to ensure that the integration does not compromise the functionality of either system. The challenge is to design a system where blockchain supports AI's dynamic learning and updating capabilities without hindering its efficiency. | [34] |

The integration of blockchain and AI (Table 5) offers significant potential benefits but also presents considerable challenges. Addressing these challenges requires ongoing research, technological innovation, and perhaps most importantly, collaborative efforts between experts in both fields to devise solutions that harness the strengths of both technologies without undue compromise.

## 7. CONCLUSION

Blockchain's attributes of decentralization, immutability, and transparency make it a transformative tool for enhancing AI transparency and trustworthiness. By creating a clear, auditable trail of AI decisions and data usage, blockchain has the potential to demystify complex AI operations, ensuring that stakeholders, regulators, and end-users can verify and understand AI-driven processes. This enhanced accountability is particularly crucial in high-stakes sectors such as healthcare, finance, and autonomous systems, where explainability and ethical considerations play a pivotal role in decision-making. However, the integration of blockchain and AI is not without challenges. Scalability issues, stemming from blockchain's transaction speed and storage limitations, must be addressed to handle the vast volumes of AI-generated data efficiently. Additionally, the computational complexity of merging AI inference mechanisms with blockchain's immutable nature necessitates the development of optimized consensus algorithms, hybrid architectures, and privacy-preserving techniques to ensure both efficacy and efficiency.

**Call to Action & Future Directions**

Realizing the full potential of blockchain in enhancing AI transparency requires sustained interdisciplinary research and industry collaboration. Moving forward, efforts should focus on:

a. Developing Scalable Blockchain-AI Frameworks – Advancing layer-2 solutions, sharding, and hybrid architectures to address blockchain's throughput and storage limitations.

b. Enhancing AI Explainability via Blockchain – Integrating explainability tools (SHAP, LIME, counterfactual reasoning) directly into blockchain-verified AI models.

c. Exploring Regulatory Alignment – Establishing frameworks that ensure compliance with data privacy laws (GDPR, AI Act) while maintaining transparency.

d. Validating Blockchain-AI Applications – Conducting real-world case studies in critical sectors to measure feasibility, security, and efficiency.

The urgency of blockchain-AI research cannot be overstated. As AI continues to shape industries and societies, ensuring its transparency, fairness, and accountability will be imperative for ethical AI adoption. By leveraging blockchain's capabilities, we have the opportunity to establish a new paradigm for AI governance—one that fosters trust, mitigates risks, and ensures responsible AI deployment at scale.

While there are hurdles to overcome, the potential benefits of integrating blockchain with AI are transformative. With strategic innovation, policy alignment, and industry adoption, blockchain-enhanced AI can redefine trust and security in automated decision-making, paving the way for a more transparent, explainable, and ethically aligned future.

**Funding:** No funding

**Institutional Review Board Statement:** Not Applicable

**Data Availability Statement:** Not Applicable

**Acknowledgement:** The authors would like to thank everyone, just everyone!

**Conflict of Interest:** The authors declare no conflict of interest.

**Authors**

**Afroja Akther** is a Joint Director at Bangladesh Bank (The Central Bank of Bangladesh) and is currently pursuing a Master of Science in Information Technology (MSIT) at Emporia State University. She specializes in Information Technology, Blockchain, Finance, and Cybersecurity. With extensive experience in financial technology (FinTech) and regulatory frameworks, she plays a key role in integrating emerging technologies into the banking sector. Her expertise includes blockchain-based financial solutions, cybersecurity risk management, and IT governance. Afroja is committed to enhancing digital financial security and driving innovation in the banking industry.

**Ayesha Arobee** is currently pursuing a Master of Business Administration (MBA) and a Master of Science in Information Technology (MSIT) at Emporia State University. Her research interests span Artificial Intelligence (AI), Information Technology (IT), Cybersecurity, and Blockchain, with a focus on AI-driven security frameworks, decentralized technologies, and IT governance. She is passionate about exploring innovative solutions that enhance transparency, security, and efficiency in digital systems.

**Abdullah Al Adnan** is currently pursuing a Master of Science in Information Technology (MSIT) at Emporia State University. His research interests focus on Information Technology and Cybersecurity, with a particular emphasis on network security, data protection, and emerging threats in digital infrastructure. He is passionate about exploring innovative cybersecurity solutions to enhance the security and resilience of IT systems.

**Omum Auyon** is currently pursuing a Master of Science in Information Technology (MSIT) at Emporia State University. His research interests lie in Blockchain, Artificial Intelligence (AI), and Cybersecurity, focusing on the integration of decentralized technologies, AI-driven security frameworks, and advanced threat detection systems. He is passionate about exploring innovative solutions to enhance transparency, security, and efficiency in digital systems.

**Johirul Islam** is currently pursuing a Master of Science in Information Technology (MSIT) at Emporia State University. His research interests focus on Artificial Intelligence (AI), Communication, and Cybersecurity, with a particular emphasis on AI-driven communication systems, secure data transmission, and emerging cybersecurity threats. He is passionate about exploring cutting-edge technologies to enhance secure and efficient digital communication frameworks.

**Farhad Akter** is currently pursuing a Master of Science in Information Technology (MSIT) at Emporia State University. His research interests focus on Artificial Intelligence (AI), Information Technology, and Cybersecurity, with an emphasis on AI-driven security solutions, IT infrastructure management, and emerging cybersecurity challenges. He is passionate about leveraging advanced technologies to enhance digital security, automation, and efficiency in IT systems.